\documentclass[aps,prb,reprint,showpacs,superscriptaddress,groupedaddress]{revtex4-1}
\usepackage{graphicx}
\usepackage{amsmath}
\usepackage{amssymb}
\usepackage{dcolumn}
\usepackage{dsfont}
\usepackage{latexsym}
\usepackage{rotating}
\usepackage{color}
\usepackage{latexsym}
\usepackage{natbib}
\usepackage{bbm}
\usepackage{subfigure}
\usepackage{float}
\usepackage{epsfig}
\usepackage{psfrag}
\usepackage{bm}
\usepackage{amsthm}
\usepackage{eucal}
\usepackage{mathrsfs}
\usepackage{braket}
\usepackage{psfrag}
\usepackage{enumerate}
\usepackage{longtable}
\usepackage{subfigure}
\usepackage{bm}
\usepackage{hyperref}
\newcommand{\be}{\begin{equation}}
\newcommand{\ee}{\end{equation}}
\newcommand{\bn}{\begin{eqnarray}}
\newcommand{\en}{\end{eqnarray}}

\usepackage{color} 

\usepackage{hyperref}
\hypersetup{
colorlinks=true,final=true,
        linkcolor=red,
        citecolor=blue,
        filecolor=blue,
        urlcolor=blue,
}

\begin{document}

\title{Feasibility of a metamagnetic transition in correlated systems} 

\author{Swagata Acharya$^{1}$}\email{acharya.swagata@phy.iitkgp.ernet.in}
\author{Amal Medhi$^{2}$}
\author{N. S. Vidhyadhiraja$^{4}$}
\author{A. Taraphder$^{1,3}$}\email{arghya@phy.iitkgp.ernet.in}
\affiliation{$^{1}$Department of Physics, Indian Institute of Technology,
Kharagpur, Kharagpur 721302, India.}
\affiliation{$^{2}$ School of Physics, Indian Institute of Science Education
and Research, Thiruvananthapuram 695016, India}
\affiliation{$^{3}$Centre for Theoretical Studies, Indian Institute of
Technology Kharagpur, Kharagpur 721302, India.}
\affiliation{$^{4}$Theoretical Sciences Unit, Jawaharlal Nehru Centre For
Advanced Scientific Research, Bangalore 560064, India.}

\begin{abstract}
The long-standing issue of the competition between the magnetic field and the Kondo effect, 
favoring, respectively, triplet and singlet ground states  
is addressed using a cluster slave-rotor mean field theory for the Hubbard model and 
its spin-correlated, spin-frustrated extensions in 2 dimension. The metamagnetic jump is 
established and compared with earlier results of dynamical mean-field theory. The present 
approach also reproduces the emergent super-exchange energy scale in the insulating side. 
A scaling is found for the critical Zeeman field in terms of the intrinsic coherence scale 
just below the metal-insulator transition where the critical spin fluctuations are soft. The 
conditions for metamagnetism to appear at a reasonable field are also underlined. The Gutzwiller 
analysis on the 2D Hubbard model and a quantum Monte Carlo calculation on the Heisenberg spin system 
are performed to check the limiting cases of the cluster slave-rotor results for 
the Hubbard model. Low-field scaling features for magnetization are discussed.
\end{abstract}

\maketitle

\section{Introduction}
The question of metamagnetism in strongly correlated systems has been a long-studied subject.
In the absence of applied magnetic field, Hubbard model shows a metal-insulator 
transition (MIT) driven by local correlation or doping~\cite{imada,georges}. 
The physics close to the MIT is likened to the formation and subsequent quenching of 
the moments in the Kondo impurity model~\cite{hewson}. The connection between the two is 
borne out in the dynamical mean-field theory where the correlated lattice model is mapped on 
to an impurity model, exact in infinite dimension. The quenching of the local 
spin fluctuations leads to Fermi liquid (FL) behavior at low temperatures, 
as shown in \cite{rozen}. This self-consistent emergence of a low energy 
coherence is typical of strongly correlated systems. Close to a Mott transition, 
such systems show a residual AF exchange between local moments. How an external magnetic field 
interacts with the moments, especially in the 
correlated metallic regime where spin fluctuations are still extant, is 
a question of considerable interest. The experimental observation of metamagnetic transition (MMT) 
in single crystal of bilayer perovskite metal Sr$_3$Ru$_2$O$_7$~\cite{perry,millis1}, 
an insulating magnet BiMn$_2$O$_5$~\cite{jwkim,jeon}, multiferroic 
hexagonal insulator HoMnO$_3$~\cite{choi}, and heavy fermions 
like MnSi~\cite{thessieu}, CeRu$_2$Si$_2$~\cite{mignot} rekindled the interest recently.

Gutzwiller approximation-based approaches~\cite{vollhardt,spalek} motivated by the question of MMT in 
He$^3$ showed that there indeed is an MMT beyond a critical correlation (lower than 
the critical Hubbard-$U$ for MIT). The presence of an MMT is, however, contradicted by 
Weigers~\cite{weigers} et al. who found a smooth variation of magnetization with 
applied field, in tune with Stoner's~\cite{anderson,ma,monod,levin} approach.
However, Stoner's theory, essentially a high temperature approach, underestimates
the local correlation and misses the competition between the local moments and 
the magnetic field. Gutzwiller approximation, on the other hand~\cite{vollhardt}, 
is incapable of describing the ground state of the correlated metallic phase 
properly. However, it is a theory for the ground state only and neglects spin correlations entirely. 
Theoretical attempts~\cite{georges,laloux,parihari} have recently been brought to bear 
upon this problem recently to understand it using more powerful techniques.
The emergence of dynamical mean-field theory (DMFT) has seen a major paradigm shift in 
the study of strongly correlated systems. These calculations reveal the presence of MMT in a 
half-filled Hubbard model. As DMFT captures the dynamics close to the transition in 
great detail~\cite{georges} and treats the local spin fluctuations better, the corresponding 
low energy scale naturally emerges. In the weakly correlated metal, a smooth transition is 
observed from an unpolarized metal to a polarized band insulator. In the strong 
coupling limit (close to  $U_{c}$), the phenomenon is distinctly different 
though $-$ showing a metamagnetic jump which drives the system from 
a strongly correlated metal to a field-driven band insulator. DMFT, however, 
could be numerically more intensive, depending on the choice of the 
impurity solver: moreover, the cluster extension of DMFT (C-DMFT) and the retrieval 
of spin-fluctuation energy scale in the insulating side are fairly demanding tasks.

While single-site DMFT and C-DMFT are perhaps some of the most efficient 
techniques for correlated systems, they have their own frailties in describing 
the insulating state with magnetic order. Incorporating spin correlations in these 
approaches is non-trivial. In addition, strongly correlated systems often 
come up with situations where spin and charge of an electron appear to behave distinctly 
and their responses to external probes are quite disparate. Such situations  
are missed in DMFT-based theories.  
A natural separation of these two distinct degrees is the key to slave-rotor (SR) 
approach. The Hilbert space of the physical electron is decoupled into the so-called 
``chargon'' (conjugate to rotor) and ``spinon''  spaces and 
the unphysical states are eliminated via  
local constraints. The strongly correlated problem then maps 
on to interacting slave particles self-consistently coupled to a 
gauge field. The gauge fluctuations, being weak, provide a framework 
for studying the Mott-Hubbard physics at intermediate to large 
coupling~\cite{florens, paramekanti} in a straightforward manner. 

The present work uses slave-rotor mean-field (SRMF) theory to investigate the MMT
within the Hubbard-Heisenberg model. In what follows we consider only the Zeeman 
field, as the orbital contribution is much weaker. We not only find the MMT but 
also address the question: \textit {why does metamagnetism (MM), arising out of 
competition between Zeeman field and Kondo fluctuation, remain so elusive 
experimentally}? We predict a scaling behavior for the critical Zeeman field in 
terms of the Kondo scale, in the strongly coupled metallic regime. A possible 
experimental realisation of the transition in Mott-Hubbard systems is prescribed as well. 
We show that tuning the system to strong coupling (via tensile strain, for example) can 
act as a precursor for field-driven MMT. We identify the regimes for the observation of MM 
at the emergent `spin-exchange' energy scale in the insulating side. Possible 
experimental realization of MM in real materials or optical 
lattices, out of correlation and external driving field, is the primary focus 
of the present work. Using extensive qualitative and quantitative arguments and 
a standard semi-analytical technique, we analyze the possible emergence of 
MM in correlated electronic systems. Indeed, there are various other slave-particle 
mean-field techniques used in the context of the Hubbard model~\cite{kotliar,lee}. As 
we see below, the SRMF theory gives good results at a nominal numerical cost 
particularly in the strong coupling limit, where MM is most likely to be observed.

\section{Model and Formalism}

We consider the $t-t'-U-J$ model on a square lattice in the presence of 
external magnetic field. This is a general correlated model without 
particle-hole symmetry at half-filling, and AF spin-exchange built 
in. The various models studied below are different limiting cases of this,
discussed as we go along.

\begin{align}
H=&-\sum_{i,j,\sigma}t_{ij}c_{i\sigma}^{\dag}c_{j\sigma}+
U\sum_{i}n_{i\uparrow}n_{i\downarrow}
\nonumber\\&+J\sum_{i,j}{\bf S}_{i}{.}{\bf S}_{j}-h\sum_{\sigma}\sigma 
n_{i\sigma}.
\end{align}

\noindent where $t_{ij}=t,\, t'$ are the nearest and next nearest neighbor 
hopping amplitudes respectively. $c_{i\sigma}^{\dag}(c_{i\sigma})$ is electron
creation (annihilation) operator at a given site. $n_{i\uparrow} (n_{i\downarrow})$ is the
density operator for the up (down) spin. $J (>0)$ introduces antiferromagnetic (AF) spin 
exchange between spins at neighboring sites, while $h$ is the external Zeeman 
field. In terms of rotor and spinon operators, this model can be written as 
(see Florens, et al.~\cite{florens} for SRMF formulation)   
\begin{align}
H_{SR}=&-\sum_{i,j,\sigma}t_{ij}f_{i\sigma}^{\dag}f_{j\sigma}e^{-i\theta_{i}}e^{i\theta_{j}}+
U/{2}\sum_{i}n_{i}^{\theta}(n_{i}^{\theta}-1)\nonumber\\&+J\sum_{i,j}{\bf S}^{f}_{i}
{\bf S}^{f}_{j}
\end{align} 
\noindent $f_{i\sigma}^{\dag}$ is the spinon creation operator, and the
rotor creation (annihilation) operator is $e^{i\theta_{i}}$ ($e^{-i\theta_{i}}$). $n_{i}^{\theta}$
is chargon density operator and ${\bf S}^{f}_{i}$ is $f_{i\alpha}^{\dag}{\bf \sigma}_{\alpha\beta}f_{j\beta}$.
\noindent In the SRMF approximation the local constraint is relegated to a 
global constraint satisfied on the average.  
\begin{equation}
\langle n_{i}^{\theta}\rangle +\langle n^{f}_{i,\uparrow}\rangle+\langle 
n^{f}_{i,\downarrow}\rangle =1.
 \end{equation}
\noindent  $\langle n_{i}^{\theta}\rangle$ and $\langle n_{i}^{f}\rangle$ are the 
average chargon and average spinon density respectively. Following straightforward algebra, 
the Hamiltonian Eq.2 decouples into two coupled Hamiltonians solved self-consistently under 
the saddle-point approximation.  
\begin{align}
H_f=&-\sum_{i,j,\sigma}t_{ij}B_{ij}f_{i\sigma}^{\dag}f_{j\sigma}+J\sum_{i,j}{\bf S}^{f}_{i}
{\bf S}^{f}_{j}-\sum_{i\sigma} (\mu_{f}+h\sigma) n_{i\sigma}^{f}
\end{align} 
\begin{align}
H_\theta=&-2\sum_{i,j,\sigma}t_{ij}\chi_{ij}e^{-i\theta_{i}}e^{i\theta_{j}}+
U/2\sum_{i}(n_{i}^{\theta})^{2}\nonumber\\&-\mu_{\theta}\sum_{i\sigma}n_{i}^{\theta}
\end{align}
\noindent where $B_{ij}=\langle e^{-i\theta_{i}}e^{i\theta_{j}}
\rangle_{\theta}$ and $\chi_{ij}=\langle f_{i\sigma}^{\dag}f_{j\sigma}
\rangle_{f}.\,\, \mu_{f},\,\mu_{\theta}$ are Lagrange multipliers for the
number constraint: two multipliers are generally used for convenience to control 
$\langle n_{i}^{\theta}\rangle$ and $\langle n_{i}^{f}\rangle$ separately 
while still satisfying Eqn.(3). In the presence of spin density wave (SDW) with 
commensurate ordering wave vector ${\bf Q}$=($\pi,\pi$), the mean-field spinon Hamiltonian is 
\begin{align}
H_{f}^{MF}=&\sum_{{\bf k}\sigma}\epsilon_{\bf k,\sigma}f^{\dagger}_{{\bf k}\sigma}
f_{{\bf k}\sigma}-2Jm\sum_{\bf k}(f^{\dagger}_{{\bf k}\uparrow}f_{{\bf k+Q}\uparrow}-
f^{\dagger}_{{\bf k}\downarrow}f_{{\bf k+Q}\downarrow})
\end{align}
\noindent where, 
\begin{align}
\epsilon_{k\sigma}=&-2(tB+3J\chi/4)(cosk_{x}+cosk_{y})
\nonumber\\&-4t^{'}B^{'}cosk_{x}\,cosk_{y}-\mu_{f}-h\sigma
\end{align}
\noindent and $\lambda^{\pm}_{k\sigma}=\frac{1}{2}(\epsilon_{k\sigma}+
\epsilon_{k+Q\sigma})+\frac{1}{2}E_{k}$\, ;\,  
$E_{k}=[(\epsilon_{k\sigma}-\epsilon_{k+Q\sigma})^{2}+
(4Jm)^{2}]^{\frac{1}{2}}$. At half-filling, the corresponding self-consistency equations for 
the spinon sector are easily obtained and the magnetization is  
\begin{align}
M=&\frac{1}{2}\sum_{k,\sigma}\sigma(n_{F}(-\lambda^{-}_{k\sigma})+
n_{F}(-\lambda^{+}_{k\sigma}))
\end{align}
\noindent $\chi, \, \chi^{'}$ are, respectively, spinon kinetic energies for nearest and
next nearest neighbor hoppings and $m$ is staggered magnetization.
\subsection{Single site theory}
\noindent The simplest version of SRMFT involves decoupling at the single site level,
neglecting intersite correlations in the mean-field. The single-site mean-field 
Hamiltonian for the rotor sector is 
\begin{align}
H_{\theta}=-8(t\chi+t^{'}\chi^{'})\phi(e^{-i\theta}+e^{i\theta})+U/2(n^{\theta})^{2}-\mu_{\theta}n^{\theta}
\end{align}
In the present approximation $B=B^{'}=\phi^{2}$. This rotor kinetic energy acts as 
the order parameter-quasiparticle (QP) weight ($Z$) for the fermions. If 
$\phi^{2}$ vanishes, the system is driven into a Mott insulating state with no 
charge fluctuations. In that case, clearly, once $\phi$ is zero the effect of Coulomb 
correlation on the spinon part also vanishes. It is, therefore, necessary to take
account of the intersite correlations to study magnetic interactions in the strongly 
correlated regime.  

\subsection{Two site (cluster) theory}
We extend the theory to the bond (cluster) approximation in view of 
the shortcomings of the single-site theory. Here $B^{'}$ should be different 
from $B$ - the nearest and next nearest neighbour correlations in 
charge sector are different. The rotor Hamiltonian for the two-site cluster is 
\begin{align}
H_{\theta}=&-2t\chi(e^{-i\theta_{1}}e^{i\theta_{2}}+h.c.)
\nonumber\\&+(6t\chi+8t^{'}\chi^{'})\phi(e^{-i\theta_{1}}+e^{-i\theta_{2}}+h.c.)
\nonumber\\&+U/2(n^{\theta}_{1})^{2}+U/2(n^{\theta}_{2})^{2}-\mu_{\theta}
(n^{\theta}_{1}+n^{\theta}_{2})
\end{align}

\noindent This Hamiltonian is again diagonalized numerically in the basis 
$\ket{n_{1}^{\theta},n_{2}^{\theta}}$, where $B^{'}$=$\phi^{2}$ 
with $\phi=\langle e^{\pm i\theta}\rangle$ and $B$=$\langle 
e^{-i\theta_{1}}e^{i\theta_{2}}\rangle$. When $\phi$ goes to $0$ 
(i.e., the insulating phase), the nearest neighbour inter-site correlation, 
absent in the single-site case, 
could assume a non-vanishing value in the cluster approximation. The first 
term in equation (10) gives a finite rotor kinetic energy ($\sim t^{2}/U$), which, in turn, 
affects the spinon sector and makes the otherwise sterile $\phi=0$ phase 
interesting. A bond approximation approach, therefore, is capable of recovering 
the inter-site spin correlation scale. 

\vspace{0.5em}
\section{Results and discussions}

To begin with, we briefly discuss our results on the Hubbard model at half-filling 
at $T=0$ (all energies are given in units of $t$) and identify the Mott transition 
in SRMF. Results on MMT in this limit exists in DMFT~\cite{georges,parihari} and a 
comparison of the same using SRMF theory is therefore in order. No result on MMT is 
available on this model using SRMF theory (or any other slave-particle methods) so 
far. We begin by studying single-site and cluster approach for the Hubbard 
Hamiltonian with and without field respectively. We also observe how the spin-spin 
exchange interferes with correlation and how the spin fluctuations are affected in 
the proximity of MIT. 

\subsection{Hubbard ($t-U$) model on a square lattice: MIT and metamagnetism} 

\begin{figure}
\centering
\includegraphics[angle=0,width=0.95\columnwidth]{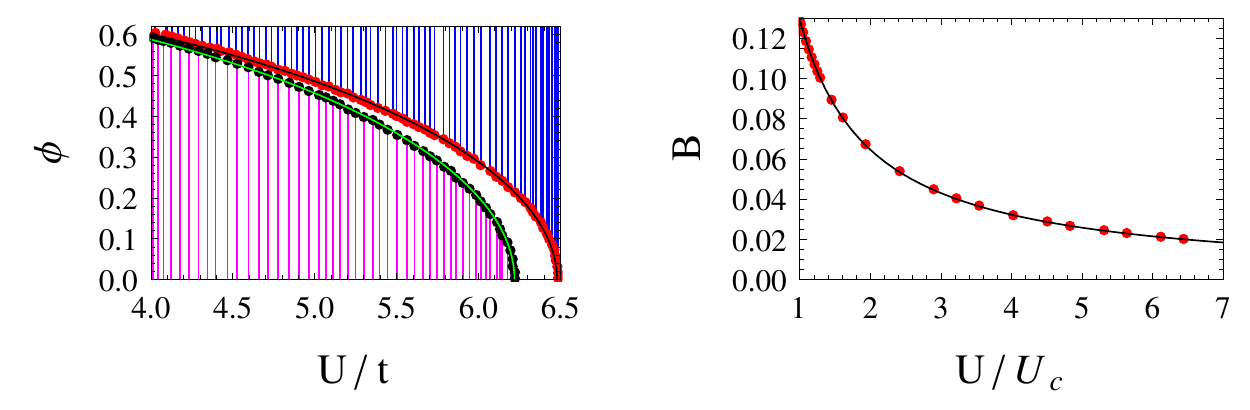}
\caption{(Color Online) Left figure: $\phi$ is plotted against $U/t$ for single-site (red dots 
are data points and the black continuous curve is a fit to Brinkman-Rice picture) and   
cluster approximation (black dots are data points and the continuous green line is the 
Brinkman-Rice fit). The critical value of $U/t$ for MIT is, 6.483 and 6.214 in single-site and 
cluster analysis respectively. 
Right figure: The inter-site correlation from the cluster analysis in the insulating phase 
(red dots) and the fit to $t^{2}/U$ (black curve).}
\label{fig1}
\end{figure}

It is well-known that in the $t-U$ model the local correlation drives the 
paramagnetic metal to a non-magnetic insulator in a Brinkman-Rice like transition 
(the order parameter $\phi$ smoothly going to zero (Fig.~\ref{fig1}, left)). The divergence 
of the effective mass is signalled by vanishing of quasi-particle weight
($Z=\phi^{2}$) at the critical point. Although the nature of the insulating or metallic 
phase remains non-magnetic (unless we resort to a two-sublattice formalism similar to that of DMFT), 
irrespective of site or cluster analysis in the SR calculations, 
we note a different critical value of $U/t$ for MIT in the site and cluster SRMF 
approach. As spatial correlations and non-local phase fluctuations are  accounted for in 
the cluster, the critical $U/t$ for MIT in a cluster approach is expected 
to be lower. The parameter ($B$) quantifying non-local fluctuations remains finite in 
the putative insulating phase and approaches zero in the $U/t\rightarrow \infty$ limit. 
For single-site analysis, we find $U_{c,site}= 6.483t$ while a two-site 
cluster gives a critical $U_{c,bond}= 6.214t$ for MIT. As far as the reduction in $U_{c}$ is 
concerned, $B$ plays much the same role as a non-local spin-spin correlation. In the 
insulating phase, $B$ is non-zero (Fig.~\ref{fig1}, right) and effective hopping remains 
finite, i.e., the effective mass does not diverge at MIT. In the cluster extension of the 
theory, therefore, spinons have a Fermi surface with finite Luttinger 
volume~\cite{anderson1} even in the insulating phase. 
\vspace{0.4em}
\vspace{-1.0em}
\vspace{0.2cm}

We note that the value $U_{c,cluster}$ for MIT that we find is in good agreement with 
the C-DMFT prediction~\cite{park} of $U_{c2}=6.050t$~\cite{commentUc}. However, the AF 
nature of the insulating ground state of the half-filled Hubbard model remains beyond 
the scope of SRMF analysis. The paramagnetic insulator in the SRMF approximation is 
an artefact of the assumption that the correlation acts in the rotor sector only, i.e., 
on the charge degrees alone; spins are free, having only a renormalized 
effective mass. There is no a priori reason, therefore, why charge ordering would 
lead to spin ordering. As a consequence the SRMF approach works better for systems 
with strong magnetic frustration, where a spin liquid insulator is likely.
Both the single site and cluster Mott transitions, have a Brinkman-Rice nature. The quasiparticle 
weight goes to to zero at critical Hubbard $U$ ($U_{c}$) in a continuous fashion, and $Z$ scales
perfectly with $\sim 1-(U/U_{c})^{2}$ (Fig.~\ref{fig1}-right panel). 

\vspace{-0.5cm}
\begin{figure}[ht!]
\centering
\subfigure[]{\label{f:C11}\epsfig{file=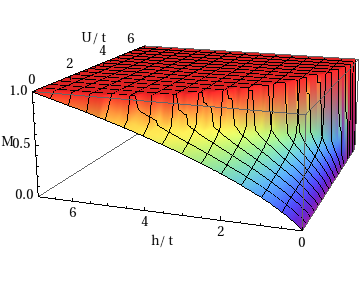,trim=0in 0in 0in 0.0in,
clip=true,width=0.48\linewidth}}\hspace{-0.0\linewidth}
\subfigure[]{\label{f:C21}\epsfig{file=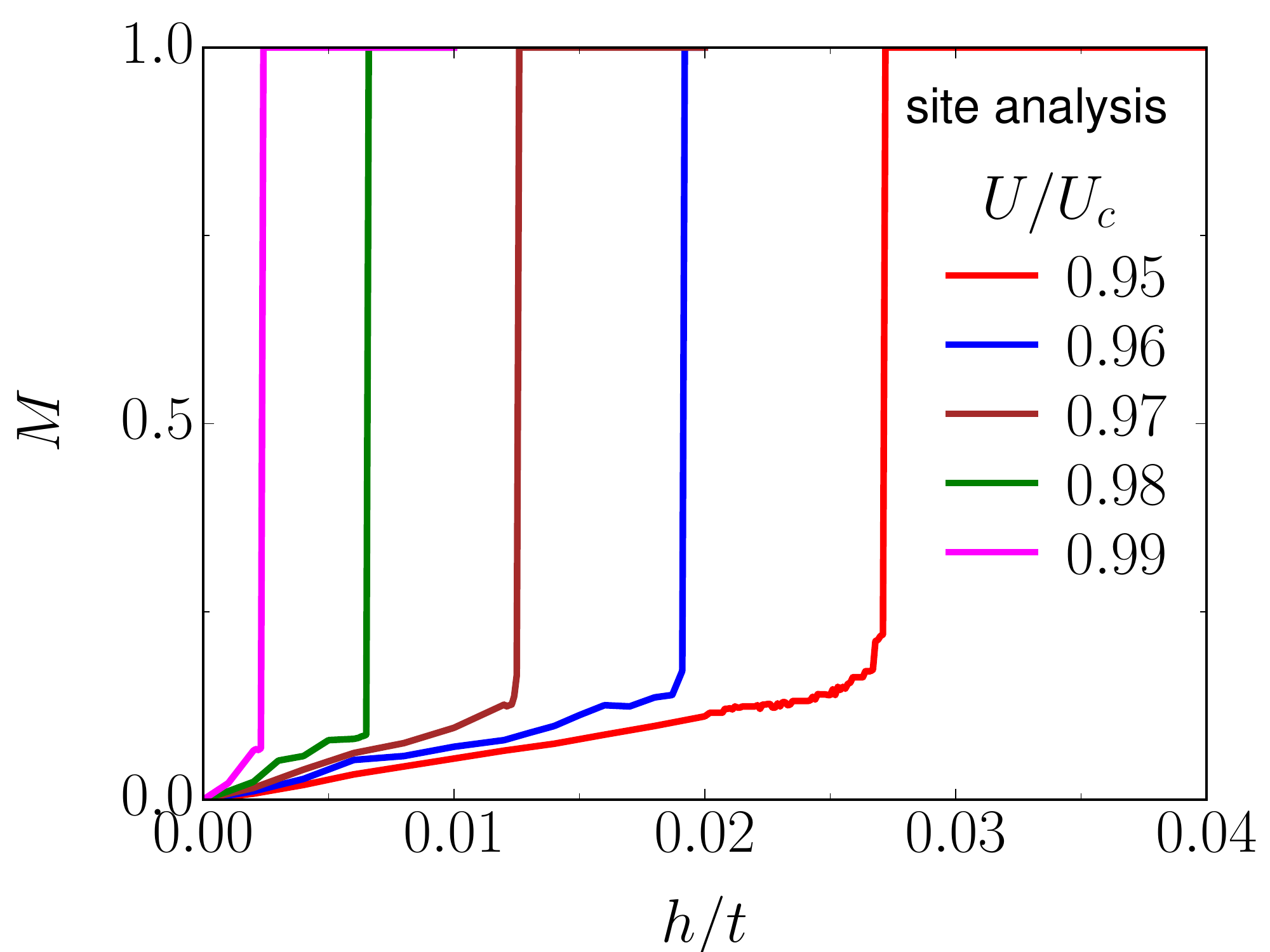,trim=0in 0in 0in 0.0in,
clip=true,width=0.48\linewidth}}\hspace{-0.0\linewidth}\\
\subfigure[]{\label{f:C21}\epsfig{file=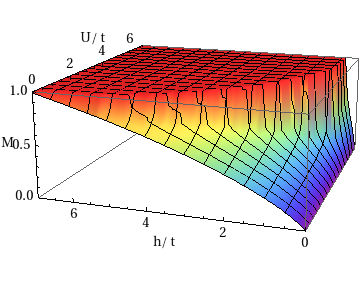,trim=0in 0in 0in 0.0in,
clip=true,width=0.48\linewidth}}\hspace{-0.0\linewidth}
\subfigure[]{\label{f:C21}\epsfig{file=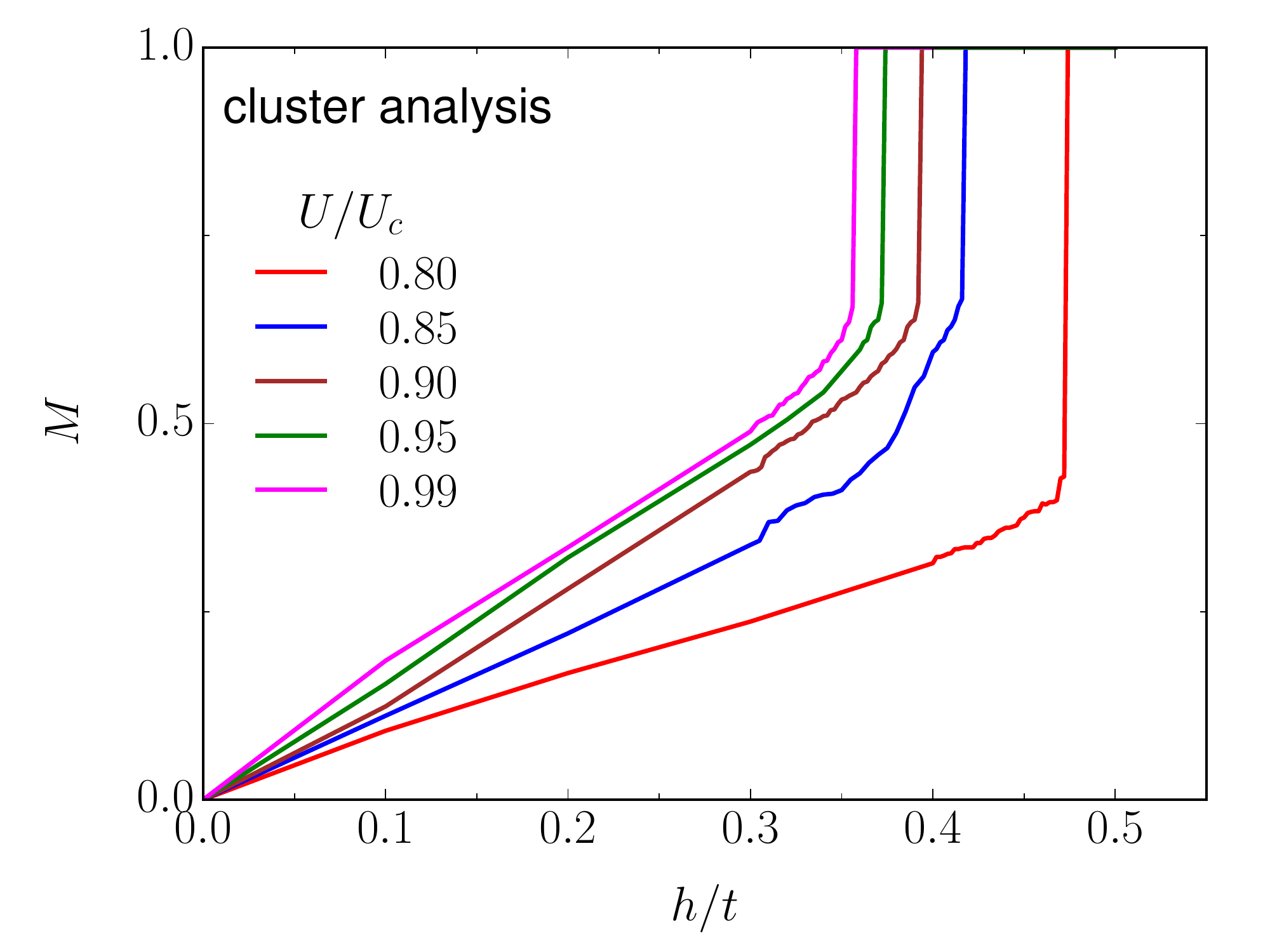,trim=0in 0in 0in 0.0in,
clip=true,width=0.48\linewidth}}\hspace{-0.0\linewidth}
\caption{Variation of magnetization showing the MMT in site [(a), (b)] and cluster [(c), (d)]  
approximation. $M-h/t$ plots close to the transition region of site and cluster theory 
are separately shown in (b) and (d) respectively.}
\label{fig2}
\end{figure}

\vspace{0.5cm}
\subsubsection{Nature of MMT: Single-site analysis}

How the system reacts to an external magnetic field, for the whole range 
of $U$ (up to U$_{c,site}$), is shown in the phase diagram 
(Fig.~\ref{fig2}(a)). For any finite $U$, ferromagnetism shows a first 
order jump to its saturation value at some critical field, clearly 
an MMT, instead of a smooth enhancement to magnetic 
saturation (as predicted originally by Stoner and extended 
later by Weigers et al.~\cite{weigers} using spin fluctuation theory). This abrupt 
jump in magnetization leads to an MIT, the order parameter going to zero ($\phi=0$) in a 
highly discontinuous manner at the same instant. This is a field-driven 
first order transition, rather than correlation-driven. The field moves the 
up and down spin bands apart leading to a weakly correlated, 
polarized band insulator~\cite{parihari}. The closer one approaches the critical 
Coulomb repulsion, it becomes more susceptible to such a transition at 
a lesser field (Fig.~\ref{fig2}(b)). Physically, at such large values of 
$U$, the kinetic degrees of freedom are nearly quenched, 
effective mass is large - the strongly correlated metal is now susceptible 
to MIT. The presence of MM is confirmed for every finite $U < U_c$.

\begin{figure}
\centering
\includegraphics[angle=0,width=0.7\columnwidth]{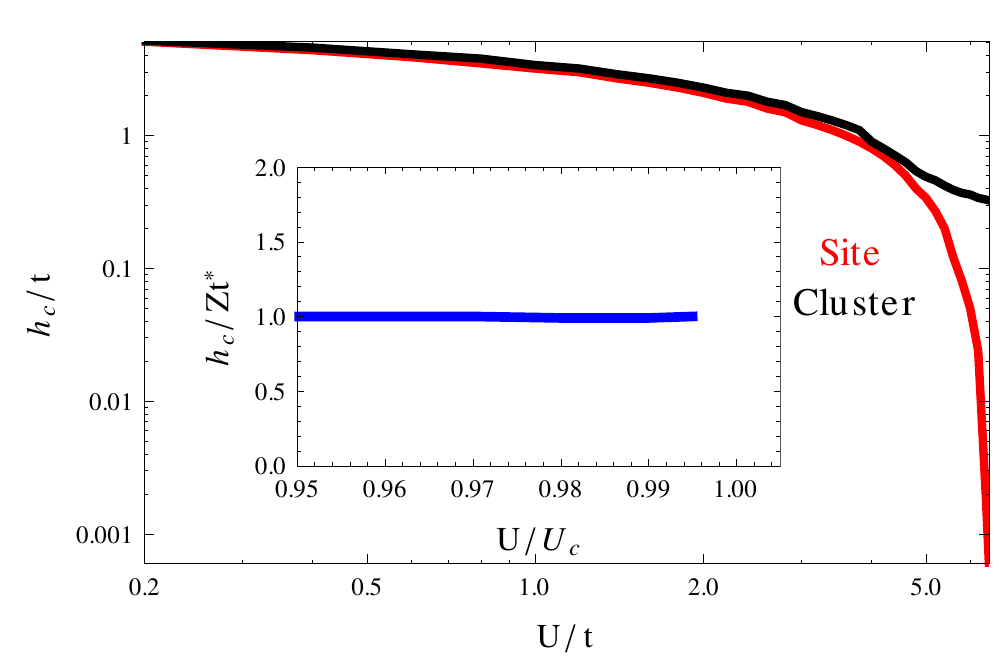}
\caption{$h_{c}/t$ in log scale against $U/t$ to highlight the huge scale variations for the critical 
numbers in the site and cluster analysis  (significantly different, not apparent from the 
phase diagram in Fig.2). Inset shows $h_{c}$, scaled with the renormalized hopping, is constant
with renormalized correlations.}
\label{fig3}
\end{figure}

\subsubsection{Nature of MMT: A two-site cluster analysis}

In the single-site theory, the insulating ground state becomes ferromagnetic for 
an infinitesimal Zeeman field. The rotor Hamiltonian being 
local, there is no magnetic exchange scale. This is a well-known  
pathology of the single-site approximation in SR. In principle, one should look for the 
competing dynamics between an 
aligning field and spin-fluctuation, recovered in the cluster version of 
the theory (Fig.~\ref{fig2}(c)). Two diagrams, site and 
cluster, are plotted (Fig.~\ref{fig2}(b),(d)) to showcase the difference between the two schemes. The MMT in the metallic state is almost similar in the two cases $-$ for any $U$ 
there is a transition. In the insulating state however, any infinitesimal $h$ 
causes saturation in magnetization ($M$) in the single-site case, while a finite critical magnetic 
field (Fig.~\ref{fig3}, main panel), representing the emergent AF spin-correlation scale 
$t^{2}/U$ (Fig.~\ref{fig1}-right panel) is required for the cluster. The putative insulating state, 
where $\phi$ becomes zero, has interesting extant dynamics via inter-site correlation $B$ within the 
cluster approximation; the rotor is still coupled to the spins and the renormalized kinetic energies
are finite in the insulating phase. In both cases, therefore, the vanishing 
of the spin-stiffness ($\chi$) signifies the metamagnetic jump: in the insulating region 
$h_c/t=0$ in single-site case while for the cluster, $h_c/t$ has to be finite (Fig.~\ref{fig3}, main panel) 
to overcome the spin-correlation scale .  

\begin{figure}
\centering
\includegraphics[angle=0,width=0.7\columnwidth]{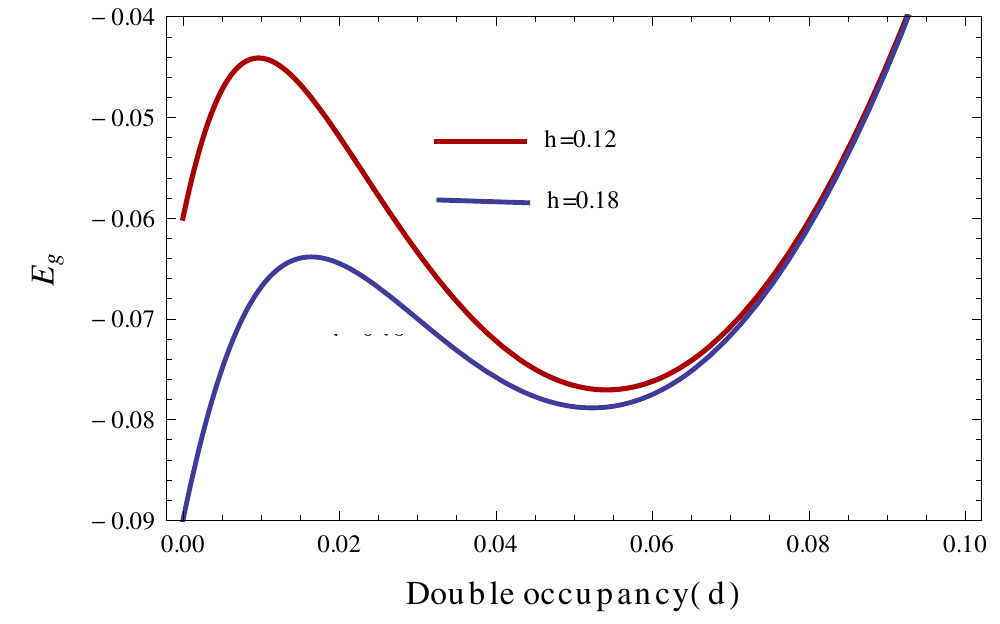}
\caption{(Color Online) Ground state energy($E_{g}$) of the Hubbard model in Gutzwiller
approximation as a function of double occupancy $d$ at $U/U_{c}=0.75$ on a square lattice 
for different $h$ (0.12, 0.18), clearly showing the first order transition.}
\label{fig4}
\end{figure}

\subsubsection{Scaling behavior for critical Zeeman field}
We note that very close to the transition, for correlated models 
$\frac{h_{c}(U)}{Zt^{*}}$ becomes independent of $U$. A strong Hubbard correlation of 
order 0.97$U_{c}$ to 0.995$U_{c}$ shows that this ratio becomes a constant
(Fig.~\ref{fig3}, inset); and it is only within this range that $h_{c}$  
has experimentally feasible values. It is interesting to note that, 
$Zt^{*}$ measures the effective renormalized bandwidth of the dispersive quasi-
particles. The strong coupling limit of the problem renormalizes this number
to a significantly smaller one, a measure of the diminishing coherence
of the quasiparticles. The $h_{c}$ in the strong coupling limit is the field required 
to destroy the coherence and favor a triplet spin state. Beyond this scale, the spin
dynamics is determined only by $h_{c}$, thereby making $\frac{h_{c}(U)}{Zt^{*}}$ universal, 
irrespective of the value of $U$. 
We find similar scaling feature for ${h_{c}}$ in higher dimensional bi-partite lattices.
The scaling should hold true for bi-partite lattices in any dimension as the hopping is renormalized 
by $\sqrt d$ factor($t^{*}=t/\sqrt d$), where $d$ is the dimension of the lattice. 
Hence the scaling is related to the competition between the coherence of the quasiparticles and 
the external agent (field) trying to destroy the coherence and is independent of the dimensionality 
of the non-interacting bath.

\subsubsection{Comparison with single-site DMFT, Gutzwiller approximation and quantum Monte Carlo (QMC)} 

In earlier studies of MMT~\cite{georges,vollhardt,parihari}, the value of the field ($h$) was limited 
below the hopping parameter $t$. In such a situation, they looked for the value of $U$ for which an MMT 
can be observed. The SRMF approach gives a critical field for MMT at any finite $U$. For  
$h<t$, we can compare our results with the previous work. For T$=0$, we compare the value of 
$U/U_{c}$ below which there is no MMT (as $h$ rises to $t$): our single-site SRMF theory gives a value 
close to 0.5 while in GA it is about 0.44~\cite{vollhardt} and in DMFT, about 
$0.61$~\cite{parihari,georges}. We also compare GA with SR MFT. Vollhardt~\cite{vollhardt} studied 
the effect of out of plane Zeeman field on the Hubbard model with a flat band using GA. While we 
choose to do the same for a square lattice dispersion, an exact analytical solution is not possible  
in this case. The value of $U_{c}$ for Hubbard model on a square lattice is about $6.451$ in GA and with 
single-site SRMFT we find it to be 6.483. The agreement between these two results motivates us to verify 
the critical value of field ($h_{c}$) that induces a metamagnetic jump for a
given $U$ within GA. This can be evaluated in two ways from GA: plot magnetization versus 
field and search for the metamagnetic point, or find the point of flipping of 
the absolute minimum in the ground state energy ($E_{g}$). The $E_{g}$ in GA is 
calculated for square lattice semi-analytically and it is a function of double 
occupancy ($d$), magnetization ($M$) and applied field ($h$). 
On minimization of energy an expression for $m$ as a function of $h$ 
and $d$ is obtained. Putting it back in $E_{g}$, the minimum is located numerically. 
The plot of $E_{g}$ against $d$ shows two minima, one at zero and the other at a finite 
value of $d$ (Fig.~\ref{fig4}). The minimum at non-zero $d$ remains the absolute minimum 
up to some critical field. At a certain $h_{c}$ the absolute minimum 
flips from a finite value to $d=0$ (Fig.~\ref{fig4}). For $U/U_{c}=0.75$, 
the critical value of field comes out to about 0.180. 
The value obtained from the SR analysis under the same set of 
parameters is $0.20$, in reasonably good agreement with GA. This similarity in 
the relevant scales emerging out of GA and SRMFT is expected 
as both of them work well in the strong coupling limit.
\vspace{0.5cm}
\begin{figure}[ht!]
\centering
\includegraphics[angle=0,width=0.95\columnwidth]{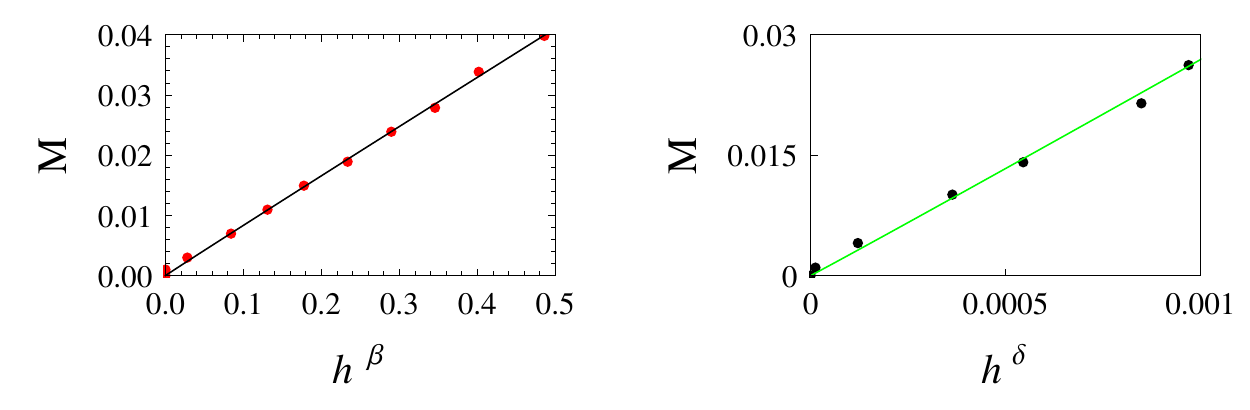}
\caption{Scaling behavior of the magnetization in a 2D square lattice for the Heisenberg model (left)
and the Hubbard model (at $U/U_{c}=0.99$, right figure). The two exponents, $\beta$ and $\delta$ are
found to be 1.02 and 0.98, where $M$ is found to scale linearly with $h^{\beta}$ and $h^{\delta}$.}
\label{fig5}
\end{figure}
 Finally we perform a QMC analysis on an AF Heisenberg spin model on a square lattice, in presence of 
perpendicular Zeeman field. On increasing the lattice sizes, we check that the $M-h$ response 
for the system remains nearly the same for $L=24,\, 32$ and 40 (where $L^{2}$ is the lattice size).
We find the exponent ($\beta$)~\cite{comm} for magnetization against field for the $L=32$ system 
and compare the number with the exponent ($\delta$) (Fig.~\ref{fig5}) in the large-$U$ limit of 
the Hubbard model. The exponents seem to agree well, which is expected since Heisenberg model
is the Large-$U$ limit of the Hubbard model at half-filling.  Though the numerically found 
exponents from the two techniques are slightly different from one (1.02 and 0.98), within our numerical
accuracy, it could well be that they are, in reality, just $1.0$ in the low field limit. 

\begin{figure}
\centering
\includegraphics[angle=0,width=0.7\columnwidth]{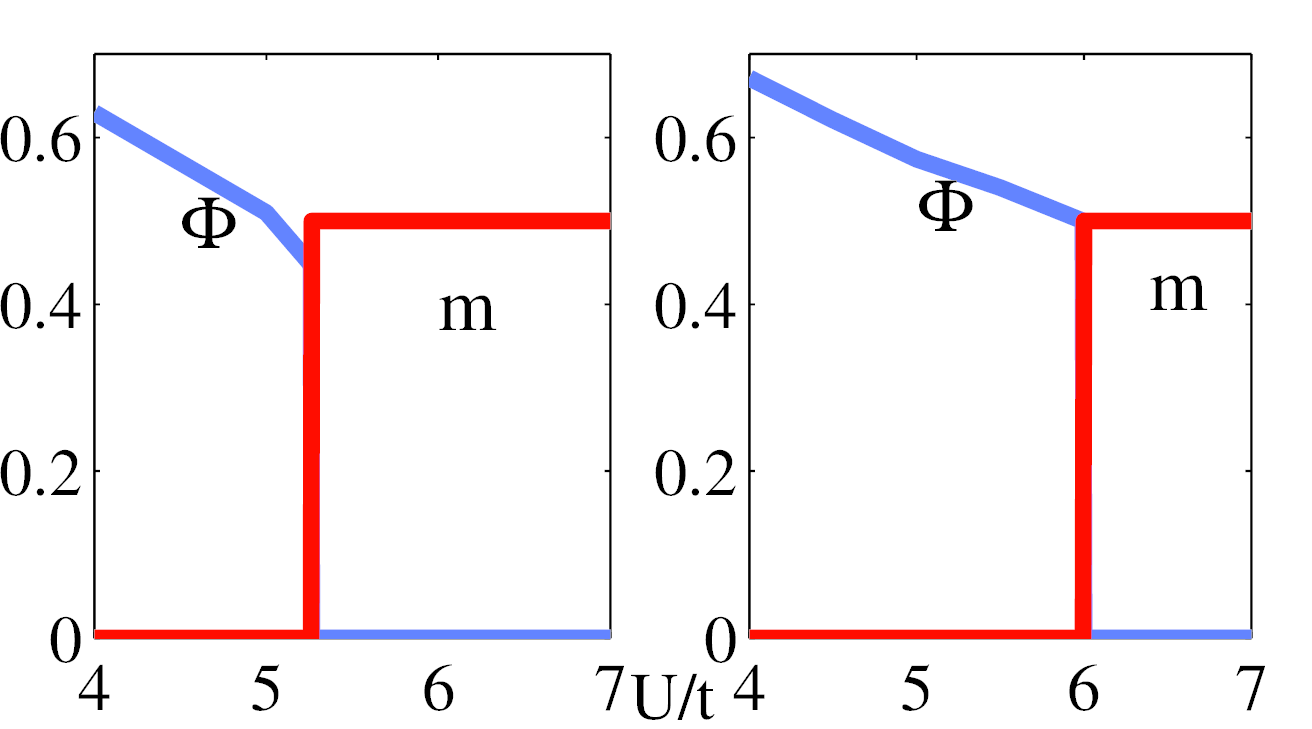}
\caption{(Color Online) Similarity of the phase diagram of 
(a) $t-t^{\prime}-U-J$ model on the square lattice and  
(b) $t-U-J$ model on a triangular lattice.  
$\phi$ is the order parameter and $m$, the staggered magnetization ($J=0.25$,
\,\, $t^{\prime}=t/2$).}
\label{fig6}
\end{figure}

\subsection{$t-U-J$ model on a square lattice}
As expected, our analysis did not produce an AF insulator in the Hubbard model
without the  explicit AF spin-exchange necessary to generate it. Therefore we use the 
$t-U-J$ model to search for a long range spin-ordered state.
It is interesting to note that an AF spin coupling $J=t/4$ 
causes a discontinuous transition from a uniform metallic state to an 
AF ordered insulating state in the case of a square 
lattice~\cite{paramekanti,florens} in the absence of field.
The metallic state  has a non-zero staggered magnetization ($m$),  
decreasing as the AF spin exchange decreases.

 In this context, we may note that the picture is quite different for a 
triangular lattice: a magnetic order (the classical Ne\'el state) and  
a sudden drop in QP weight appear at the same point (Fig.~\ref{fig6})~\cite{paramekanti}. 
The triangular lattice has no nesting, no particle-hole symmetry at half-filling
and mitigates staggered magnetization in the metallic state. The square density 
of states, on the other hand, has a logarithmic divergence at zero energy  
and is amenable to an AF order at half-filling. For the latter a non-zero 
next nearest neighbor hopping ($t^{'}$), therefore, introduces the 
necessary frustration, as opposed to the inherent geometric one in a triangular lattice. 
The same sign of $t^{'}$ as $t$ ensures a convex Fermi surface for the spinon sector.

The role of external field is to destroy the staggered magnetization and 
favor an unsaturated ferromagnetic metal and finally a saturated  
ferromagnetic insulating state in the $t-U-J$ model on a square 
lattice. However, as we discussed above the $t-t'-U-J$ model would be 
the natural choice for a  square lattice, in which, the long-range order due to 
nesting is suppressed by the frustration of the interactions. In the next section 
we discuss the MIT in the $t-t'-U-J$ model and the effect of Zeeman field on it. 
  
\subsection{$t-t^{'}-U-J$  model on a square lattice}

The single-site analysis of $t-t^{'}-U-J$ model on the square lattice shows a first order MIT at 
$U_{c}$=5.131$t$ in the absence of magnetic field. Switching over to a two-site cluster drives the 
critical correlation to a larger value 5.653$t$. In the single-site case the staggered magnetization 
saturates at the MIT and in the insulating side the spin and charge dynamics get quenched 
completely. In the cluster extension, however, the insulating state still has non-local phase 
fluctuations that lead to a finite spin stiffness and therefore the staggered magnetization 
never saturates. The finiteness of $B$ for large values of correlation implies that the 
AF order does not saturate. Hence, in the insulating side, on application of external magnetic 
field, a minimum $h=J$ is needed to destroy the AF order and revive ferromagnetism in 
the single-site theory. The metallic side of the problem in the single-site theory remains 
featureless except that there is no  AF metal now. A strong local correlation very close to MIT, 
however, competes against this frustration and produces a weakly AF metal. 

The cluster analysis reveals interesting physics in both the metallic and 
insulating regimes. While 
the MIT becomes second order in this case, the Zeeman field competes with 
the exchange energy $J$ as well as the emergent super-exchange scale in the 
insulating side. However, the MMT for all finite values of local correlation 
still survives for both site and cluster analysis. 
\begin{figure}
\centering
\includegraphics[angle=0,width=0.75\columnwidth]{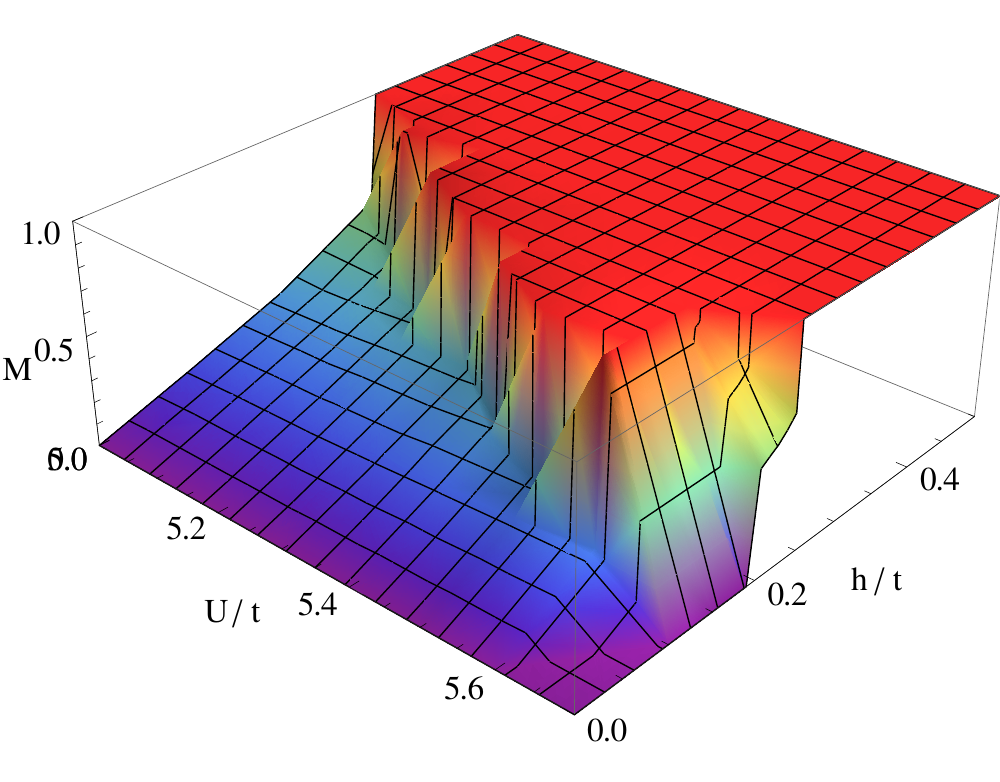}
\caption{$t-t'-U-J$ model on a square lattice ($t=1.0,\, t'=0.50,\, J=0.25$):
A close-up view of MMT in the  cluster analysis very near to the MIT. The metallic side is 
qualitatively similar to Fig.~\ref{fig2}, while the insulating side has two distinct 
jumps (see text).}  
\label{fig7}
\end{figure} 
The $M-U-h$ phase diagram (Fig.~\ref{fig7}) from the cluster analysis of the 
present model, zoomed around the MIT region, reveals interesting details beyond the 
critical local correlation. There are two distinct jumps in the magnetization 
for $U=5.7t > U_{c}=5.653t$. These two manifestly separate locations of the jump 
correspond to the competition between two different scales with the external Zeeman field: 
the first one is where antiferromagnetism gets suppressed and ferromagnetism shows up, and 
the second one is where the spins are ferromagnetically saturated at an energy scale  
given by the residual kinetic energy ($Bt\sim t^{2}/U$). 

\section{Experimental realizations} 
Metamagnetism has, of late, become a 
highly recurrent~\cite{perry,jwkim,jeon,thessieu, choi,mignot,millis} phenomenon. 
The perpendicular or in-plane field required for such discontinuous, super-linear
transitions in magnetization can be under 10 Tesla. All the strongly 
correlated systems mentioned earlier show MMT between 1 to 10 
Tesla fields. However, the MM in each of these is material-specific and often 
associated with some structural transitions. The role of correlation is not 
clear in most of them, while our concern is MM out of 
correlation alone where the competition between the applied field and 
the local spin fluctuation is the key in driving the non-linear 
magnetization and the consequent first order jump. 

On the long-standing issue of MM in liquid He$^3$, 
Georges and Laloux ~\cite{laloux} hold the view that liquid He$^3$ should be 
viewed as a Mott-Stoner liquid and Hubbard model with about 8\% vacancy offers 
a reasonable description of it. They predict MMT at about 26 bar in 80 Tesla 
field. Weigers~\cite{weigers}, however, did not find a metamagnetic jump in liquid 
He$^3$ up to 200 Tesla. This may indicate~\cite{laloux} that He$^3$ cannot be modelled by 
a half-filled Hubbard model. Vollhardt~\cite{vollhardt} puts liquid He$^3$ in the 
intermediate coupling regime with $U/E_{F}$ in the range less than one 
(typically 0.8 or less). As we have shown above, the fields required for MMT is 
amenable only in the strong coupling regime and therefore Vollhardt's estimates 
would imply that liquid He$^3$ is not an ideal candidate for the observation of 
correlation-driven MMT within an accessible laboratory field. 

The problem, therefore, is to find a material that can be modelled well by 
Hubbard model or any of its extended incarnations with desired range of 
parameters. In correlated systems the bare value of
$k_{B}T_{F}$ ($T_{F}$ is Fermi temperature) is nearly 1-5 eV, which ($T_{F}$) scales down to
$k_{B}T_{K}=Zt^{*}$ close to Mott MIT. Exactly at this range of parameters, as we showed, $h_{c}$
for MMT becomes $<$ 100T. The conclusion is driven by the fact that for $t=1eV$, in the site 
analysis, we get $h_{c}/t < 0.01$, which is equivalent to a field of 100 Tesla or less. 
We find that for a system with $U/t$  of order 0.99$U_{c}$, the typical value of critical Zeeman 
field, by a similar analysis, would be about 20 Tesla ($h_{c}/t < 0.002$). A strongly 
correlated system which can be reasonably modelled by single band Hubbard model, with its 
effective correlation $U/U_{c}$ tuned (by pressure, for example) somewhere between 0.97 
to 0.99, should, therefore, show an MMT within a reasonable magnetic field. 
It is also likely that cold atom systems in optical lattices should provide us 
the option of observing this in the laboratory. The scaling analysis we discuss here, 
therefore, underlines the fact that systems with narrow correlated bands or orbital selectivity 
(in multi-orbital situations, for example) can facilitate MMT at an accessible field.    
\section{Conclusions}

A ground state analysis of the Hubbard model and its spin-correlated, 
spin-frustrated versions have been performed in the present paper in search
of MM. The response to an externally applied Zeeman field in 
the SR mean field formalism shows that there indeed is a regime
of parameters where an MMT to a ferromagnetic state
occurs for all finite local Hubbard correlations. However, it is clear that to observe
MM in a laboratory field, one would need to tune the system to a very narrow range close to Mott
transition and apply a fairly large magnetic field. These conditions make the 
observation of MM in such systems so elusive. On a fundamental level, on the other hand, 
one would like to know what happens to the spin-fluctuation scale and 
the Kondo scale in a metal in the presence of a strong field favoring spin-alignment. 
Clearly, it is the sharpest Kondo resonance peak, with narrowest width (requiring close 
proximity to Mott transition), that is most sensitive to external field and leads 
to non-linear jump in magnetization and the consequent MMT.
\vspace{0.5cm}

\noindent {\bf Acknowledgement:} The authors thank Vijay B. Shenoy for discussions and 
comments on the manuscript. S.A thanks Serge Florens for useful interactions 
at the initial stage and UGC (India) for a fellowship.

\bibliographystyle{apsrev4-1}

\begin{thebibliography}{50}
\bibitem{commentUc} There are two critical correlations in DMFT 
calculations~\cite{georges}, $U_{c1}$ and $U_{c2}$. The SRMF approach has only one 
critical $U$, physically same as $U_{c2}$ of DMFT.
\bibitem{comm} We use the symbol $\beta$ to avoid confusion with the customary 
critical exponent $\frac{1}{\delta}$ at finite $T_{c}$ in 3D.
\end{thebibliography}
 
\end{document}